\documentstyle[epsf,float,twocolumn]{jpsj}

\newcommand{\lsim}
 {\ \raise.35ex\hbox{$<$}\kern-0.75em\lower.5ex\hbox{$\sim$}\ }
\def\runtitle{
Parity-Sensitive Measurements Based on
Ferromagnet/Superconductor Tunneling Junctions}
\def\runauthor{
T. {\sc Hirai}, N. {\sc Yoshida},
Y. {\sc Tanaka}, J. {\sc Inoue},
and S. {\sc Kashiwaya} }
\title{
Parity-Sensitive Measurements Based on
Ferromagnet/Superconductor Tunneling Junctions
}
\author{
Takashi Hirai$^{1,3}$, Nobukatsu Yoshida$^{1}$, Yukio Tanaka$^{1,3}$,
Junichiro Inoue$^{1}$ \\and Satoshi Kashiwaya$^{2,3}$}
\inst{
$^{1}$Department of Applied Physics, Nagoya University,
Nagoya, 464-8603\\
$^{2}$
Electrotechnical Laboratory, Tsukuba, Ibaraki 305-8568, Japan \\
$^{3}$CREST, Japan Science and Technology Corporation (JST)
}
\recdate{\today}
\abst{
A method for identifying the parity
of unconventional superconductors
based on tunneling spectroscopy is proposed.
For a model of calculation, we adopt a ferromagnet/superconductor (F/S)
junction, the tunneling current of which is spin polarized.
The tunneling conductance spectra are shown to be
quite sensitive to the
direction of the magnetization axis in the ferromagnet
only when the superconductor has odd parity.
Therefore, it is possible to
distinguish the parity of the superconductor
by performing tunneling spectroscopy in F/S junctions.
}
\kword
{tunneling conductance, triplet superconductor,
parity of the superconductor, Andreev reflection,
ferromagnet
}
\begin{document}
\sloppy
\maketitle
The discovery of superconductivity in Sr$_2$RuO$_4$,
\cite{Maeno,Rice}
aroused our interest in
triplet pairing states in metals.
To date, the spin triplet pairing states have been
proposed to be realized in UPt$_3$,
\cite{Tou}
(TMTSF)$_2$PF$_6$
\cite{Lee}
and UGe$_2$.
\cite{Saxena}
In triplet superconductors,
Cooper pairs have a rich internal degree of freedom
as compared to those in singlet ones.
One of the important quantities for identifying the
basic properties of
triplet superconductors is the $d$-vector,
which characterizes the odd parity of a Cooper pair.
It is an interesting problem to develop a novel method for distinguishing
the parity of the superconductors based on 
the dependence of the direction of the $d$-vector in tunneling spectroscopy.
\par
The tunneling effect in normal metal / insulator / unconventional
superconductor ($N/I/S$) junctions
markedly reflects the symmetry of the pair potential
of unconventional superconductors.
In particular,
the sign change in the pair potential induces a significant effect,
{\it i.e.},
a zero-bias conductance peak (ZBCP)
in tunneling experiments of high-$T_C$ superconductors.
\cite{Kashiwaya1,Wei,Wang}
Using a tunneling conductance formula in
normal metal / insulator / unconventional singlet
superconductor junctions,
the origin of the ZBCP is explained in terms of the
formation of the zero-energy Andreev bound states (ZES) at the interface of
a superconductor. \cite{Hu,Tanaka1,Kashiwaya2,Kashiwaya4}
Applying this formula to triplet superconductor junctions,
ZBCPs are also obtained
\cite{Buchholtz,Hara,Yamashiro,Honerkamp}.
Since the ZBCP is a universal phenomena expected to exist for
unconventional superconductors with
the sign change in the pair potential on the Fermi surface
regardless of the parity,
we cannot distinguish the parity of the
superconductor using only tunneling spectroscopy
in the $N/I/S$ junctions.
The tunneling effect
in ferromagnet / insulator / superconductor junctions $(F/I/S)$
has also been studied.
\cite{Jong,Zhu,Kashiwaya3,Zutic,Yoshida}
Since the retro-reflectivity of the Andreev reflection
\cite{Andreev}
is broken due to the exchange potential in the ferromagnet,
the height of ZBCP is suppressed when the superconductor is
singlet.
With regard to triplet superconductor junctions,
the situation becomes much more complex.
It is revealed that whether the magnitude of ZBCP is suppressed or not
strongly depends on 
the direction of the $d$-vector.
However, there is no systematic study to
clarify the influence of the ferromagnet
on the tunneling conductance in triplet
superconductor junctions at this stage.
To more clearly reveal the difference
between singlet superconductors and triplet ones
through tunneling conductance in $F/I/S$ junctions,
we must propose a new idea.
\par
In this study, we calculate the tunneling conductance
for ferromagnet / insulator / superconductor ($F/I/S$) junctions
with arbitrary direction of the magnetization axis.
It is clarified that
the tunneling conductance for triplet superconductors depends on
the angle between the magnetization axis and the $d$-vector,
while that for singlet ones does not
because the total spin angular momentum of the Cooper pair is zero.
Through the change in the tunneling conductance as a function of
the direction of the magnetization axis,
we can identify much more detailed features of
a triplet superconducting Cooper pair.
With this idea, we can clarified detailed features of
the pair potential of Sr$_2$RuO$_4$.
\par
%
%
We assume a two-dimensional
$F/I/S$ junction with semi-infinite double-layered structures
in the clean limit.
A flat interface is perpendicular to the $x$-axis
and is located at $x=0$. The insulator is modeled
as a delta-functional form $V(x)=H\delta (x)$, where $H$ and $\delta (x)$
are the height of the barrier potential and $\delta$ function, respectively.
The Fermi energy $E_F$ and the effective mass $m$
are assumed to be equal in both the ferromagnet
and the superconductor.
As a model of the ferromagnet, we apply the Stoner model
using the exchange potential $U$.
The magnitude of momentum in the ferromagnet for the
majority ($\bar{\uparrow}$) or minority ($\bar{\downarrow}$) spin is denoted by
$k_{\bar{\uparrow} (\bar{\downarrow} )}=\sqrt{\frac{2m}{\hbar^{2}}(E\pm U)}$.
The wave functions $\Psi({\bf r})$ are obtained by solving the
Bogoliubov-de Gennes (BdG) equation
applying the quasi-classical approximation
\cite{Bruder}
%
%
\begin{eqnarray}
\left( 
\begin{array}{cc}
\hat{H}({\bf r}) & \hat{\Delta} (\theta_S ,x)\\
\hat{\Delta}^{\dag} (\theta_S ,x) & -\hat{H}^{\dag}({\bf r})
\end{array}
\right)
\left(
\begin{array}{c}
u_{\uparrow}({\bf r})\\ u_{\downarrow}({\bf r})\\
v_{\uparrow}({\bf r})\\ v_{\downarrow}({\bf r})
\end{array}
\right) =E
\left(
\begin{array}{c}
u_{\uparrow}({\bf r})\\ u_{\downarrow}({\bf r})\\
v_{\uparrow}({\bf r})\\ v_{\downarrow}({\bf r})
\end{array}
\right) ,
\end{eqnarray}
where $E$ is the energy of the quasiparticle,
$\hat{H}({\bf r})=h_0 \hat{\bf 1}-{\bf U}({\bf r})\cdot {\bf \sigma}({\bf r})$,
$h_0 =- \frac{\hbar^{2}}{2m}\nabla^{2} +V(x)-E_F$,
${\bf U}({\bf r})=U\Theta (-x){\bf n}$,
$\hat{\bf 1}$ and ${\bf \sigma}$ are
the ${2\times 2}$ identity matrix and Pauli matrix, respectively.
The quantity $\theta_S$ denotes the direction of the motion of
quasiparticles in the superconductor.
The quantity ${\bf n}$ is the direction of the magnetization axis, and
$\Theta (x)$ is the Heaviside step function.
The indices $\uparrow, \downarrow$ denote the
up and down spin in the superconductor, respectively.
The configuration of the magnetization axis of the ferromagnet
and the $c$-axis of the superconductor is expressed by
a polar coordinate $(\theta_M ,\phi_M)$ (see Fig. 1.),
where we assume that the quantization axis of the
triplet superconductor is parallel to the $c$-axis.
The effective pair potential
$\hat{\Delta}(\theta_S ,x)= \hat{\Delta}(\theta_S )\Theta (x)$
is given by
\begin{eqnarray}
\hat{\Delta}(\theta_S )=\left(
\begin{array}{cc}
\Delta_{\uparrow \uparrow}(\theta_S ) &
\Delta_{\uparrow \downarrow}(\theta_S )\\
\Delta_{\downarrow \uparrow}(\theta_S ) &
\Delta_{\downarrow \downarrow}(\theta_S )
\end{array}
\right).
\end{eqnarray}
%
It is comprehensive to rewrite the pair potential
in the coordinate of spin space in the ferromagnet,
$\Delta^{F}_{\bar{s}\bar{s}'}(\theta_S )
=\hat{U}^{\dag}\Delta_{ss'}(\theta_S )\hat{U}$,
where the unitary operator $\hat{U}$
is given by
\begin{eqnarray}
\hat{U}=\left(
\begin{array}{cc}
\gamma_1 & -\gamma_{2}^{*}\\
\gamma_2 & \gamma_{1}^{*}
\end{array}
\right)
\mbox{ , }
\left\{
\begin{array}{l}
\gamma_1=\cos{\frac{\theta_M}{2}}e^{-i\phi_M /2} \\
\gamma_2=\sin{\frac{\theta_M}{2}}e^{i\phi_M /2}
\end{array}
\right. .
\end{eqnarray}
Here, the spin indices $\bar{s},\bar{s}'=\bar{\uparrow} ,\bar{\downarrow}$
correspond to the majority and minority spin in the ferromagnet, respectively,
and $s,s'=\uparrow ,\downarrow$.
In general,
we should consider the following four kinds of reflection processes
with arbitrary $H$ and $\theta_M$
for an electron with majority spin injection:\\
i) Andreev reflection of majority spin
($a_{\bar{\uparrow}\bar{\uparrow}}$)\\
ii) Andreev reflection of minority spin
($a_{\bar{\uparrow}\bar{\downarrow}}$)\\
iii) normal reflection of majority spin
($b_{\bar{\uparrow}\bar{\uparrow}}$) and\\
iv) normal reflection of minority spin
($b_{\bar{\uparrow}\bar{\downarrow}}$).\\
Similar reflection processes also exist for minority-spin injection.
Here, $a_{\bar{s}\bar{s}'}$ and $b_{\bar{s}\bar{s}'}$
are reflection coefficients
of the Andreev and normal reflections, respectively.
The wave function of the quasiparticle 
for majority- and minority-spin injections
is denoted by the coefficients
$a_{\bar{s}\bar{s}'}$ and $b_{\bar{s}\bar{s}'}$
for $x<0$.
The coefficients
$a_{\bar{s}\bar{s}'}$ and $b_{\bar{s}\bar{s}'}$ are
determined by solving the BdG equation
with quasi-classical approximation
under the boundary condition.
The normalized tunneling conductance for zero temperature
is given by
\cite{BTK}
\begin{eqnarray}
\sigma_{T}(eV)=\frac{
\int_{-\pi /2}^{\pi /2}d\theta_S \cos{\theta_S}
\left( \sigma_{S\bar{\uparrow}}(\theta_S)+\sigma_{S\bar{\downarrow}}(\theta_S)
\right) }{
\int_{-\pi /2}^{\pi /2}d\theta_S \cos{\theta_S}
\left( \sigma_{N\bar{\uparrow}}(\theta_S)+\sigma_{N\bar{\downarrow}}(\theta_S)
\right) }
\end{eqnarray}
\begin{eqnarray}
\sigma_{S\bar{\uparrow}}&=&1+|a_{\bar{\uparrow}\bar{\uparrow}}|^2
-|b_{\bar{\uparrow}\bar{\uparrow}}|^2
+\left( \frac{\eta_{\bar{\downarrow}}}{\eta_{\bar{\uparrow}}}
|a_{\bar{\uparrow}\bar{\downarrow}}|^2
-\frac{\eta_{\bar{\downarrow}}}{\eta_{\bar{\uparrow}}}
|b_{\bar{\uparrow}\bar{\downarrow}}|^2\right)\nonumber \\
& &\times \Theta (\theta_C -|\theta_S |)
\end{eqnarray}
\begin{eqnarray}
\sigma_{S\bar{\downarrow}}&=&\left(
1+\frac{\eta_{\bar{\uparrow}}}{\eta_{\bar{\downarrow}}}
|a_{\bar{\downarrow}\bar{\uparrow}}|^2
+|a_{\bar{\downarrow}\bar{\downarrow}}|^2
-\frac{\eta_{\bar{\uparrow}}}{\eta_{\bar{\downarrow}}}
|b_{\bar{\downarrow}\bar{\uparrow}}|^2
-|b_{\bar{\downarrow}\bar{\downarrow}}|^2 \right)\nonumber \\
& &\times \Theta (\theta_C -|\theta_S |)
\end{eqnarray}
\begin{eqnarray}
\sigma_{N\bar{\uparrow}}&=&\frac{4\eta_{\bar{\uparrow}}}
{(1+\eta_{\bar{\uparrow}})^2 +Z_{\theta_S}^2}\\
\sigma_{N\bar{\downarrow}}&=&\frac{4\eta_{\bar{\downarrow}}}
{(1+\eta_{\bar{\downarrow}})^2 +Z_{\theta_S}^2}\Theta (\theta_C -|\theta_S |),
\end{eqnarray}
with $Z_{\theta_S}=Z/\cos{\theta_S}$, $Z=2mH/\hbar^2 k_F$ and
$\eta_{\bar{\uparrow} (\bar{\downarrow} )}=
\sqrt{1\pm X \cos^2 \theta_S}$.
Here, we define the polarization parameter $X=U/E_F$.
The quantity $\sigma_{S\bar{\uparrow} (\bar{\downarrow} )}$
is the tunneling conductance for electron injection
with a majority (minority) spin in the superconducting state, and
$\sigma_{N\bar{\uparrow} (\bar{\downarrow} )}$
denotes that in the normal state.
For $|\theta_S |>\theta_C =\cos^{-1}\sqrt{X}$,
the reflected wave function with a minority spin for majority-spin injection
cannot exist as a propagating wave.
Therefore, these reflection processes
do not contribute to the tunneling conductance.
In the above equation, $\theta_M$ dependence of the tunneling conductance
appears only through pair potentials.
In general, for singlet superconductors,
since $\Delta_{\bar{s}\bar{s}'}^{F}(\theta_S )=\Delta_{ss'}(\theta_S )$
is satisfied for any $\theta_M$, 
$\sigma_{S\bar{s}}$ are independent of $\theta_M$.
Thus, the $\theta_M$ dependence of tunneling conductance never occurs.
\par
%
%
As the prototype of triplet superconductors,
we consider one of the unitary pairing states presented in Sr$_2$RuO$_4$
given by
$\Delta_{\uparrow\downarrow}(\theta_S )$$=$
$\Delta_{\downarrow\uparrow}(\theta_S )$$=$
$\Delta_0 e^{i\theta_S}$ and
$\Delta_{\uparrow\uparrow}(\theta_S )$$=$
$\Delta_{\downarrow\downarrow}(\theta_S )$$=$
$0$, where the $d$-vector is along a $c$-axis.
$\hat{\Delta}^{F}(\theta_S )$
in the unitary case is given by
\begin{eqnarray}
\hat{\Delta}^{F}(\theta_S )
&=&\left(
\begin{array}{cc}
\Delta^{F}_{\uparrow \uparrow}(\theta_S ) &
\Delta^{F}_{\uparrow \downarrow}(\theta_S )\\
\Delta^{F}_{\downarrow \uparrow}(\theta_S ) &
\Delta^{F}_{\downarrow \downarrow}(\theta_S )
\end{array}
\right) \nonumber \\
&=&\left(
\begin{array}{cc}
\sin{\theta_M} & \cos{\theta_M}\\
\cos{\theta_M} & -\sin{\theta_M}
\end{array}
\right) \Delta_0 e^{i\theta_S}.
\end{eqnarray}
In particular, for $\theta_M =0$ or $\pi$($\theta_M =\pi /2$),
the relation $a_{\bar{s}\bar{s}}=b_{\bar{s},\bar{-s}}=0$
($a_{\bar{s},\bar{-s}}=b_{\bar{s},\bar{-s}}=0$) holds.
\cite{Kashiwaya3,Yoshida}
The voltage dependence of $\sigma_{T}(eV)$ is plotted
for $Z=0$ with $X=0.999$ (see Fig. 2(a)),
where only an injected electron with a majority
spin can contribute to the tunneling conductance
in both the superconducting and normal states.
As a reference, the corresponding quantity for $X=0$
is plotted in curve d, where $\sigma_{T}(eV)=2$
is satisfied for $0<eV<\Delta_0$
due to the complete Andreev reflection.
For $\theta_M =0$ or $\pi$,
since a reflected hole has a minority spin for an electron injection
with a majority spin,
the Andreev reflection does not exist
as a propagating wave for $|\theta_S |>\theta_C \sim\pi /100$.
Consequently, $\sigma_T (eV)$ is drastically suppressed
for $0<eV<\Delta_0$.
For $\theta_M =\pi /2$,
since a reflected hole has a majority spin for an electron injection
with a majority spin,
wave function of the Andreev reflected hole is a propagating wave
for arbitrary $\theta_S$.
Then, the magnitude of $\sigma_T (eV)$ does not decrease.
\par
In Fig. 2(b),
$\theta_M$ dependence of the normalized tunneling conductance
$\sigma_T (eV=0)$ is plotted
against various magnitudes of $Z$ with $X=0.9$.
The magnitude of $\sigma_{T}(eV=0)$
is significantly influenced by $\theta_M$.
The same calculation for the $d_{x^2 -y^2}$-wave superconductor is
also plotted in curve d for $Z=1$ with $X=0.9$
as a typical example of a singlet superconductor.
The resulting $\sigma_T (eV=0)$ is independent of $\theta_M$,
since total spin angular momentum of the singlet Cooper pair is zero.
It is a unique property that
using the $\theta_M$ dependence of $\sigma_T (eV=0)$,
we can distinguish the parity of the superconductor
in the $F/I/S$ junction.
\par
Next, we observe the energy dependence of the $\sigma_T (eV)$
plotted for $Z=10$ with $X=0.9$ and $0.999$ (see Fig. 3).
The magnitude of $\sigma_T (eV)$ for $\theta_M =0$ or $\pi$
is smaller than that for $\theta_M =\pi /2$,
as in the case of Fig. 2(a) (see curves $a$,$b$ and $c$,$d$).
For all curves,
there is a change in the curvature at a certain energy $eV= E_C$.
In order to explain this property,
we consider the relationship between bound states and $\theta_C$.
When the transparency of the junction is low, {\it i.e.},
the magnitude of $Z$ is large,
$\sigma_{T}(eV)$
is expressed by the surface density of states (SDOS)
of the superconductor and the
energy levels of the bound state crucially determine the
low-energy properties of SDOS \cite{Tanaka96}.
The energy levels of the bound states in the present unitary triplet pair 
potential
are obtained by solving the following equation
\cite{Hu}
\begin{eqnarray}
1=-e^{2i\theta_S}
\frac{E+i\sqrt{\Delta_{0}^{2}-E^2}}{E-i\sqrt{\Delta_{0}^{2}-E^2}},
\end{eqnarray}
and the calculated results are plotted in the inset of Fig. 3.
The bound states exist only for
$\theta_S \geq 0$ and not for $\theta_S <0$
due to the effect of the 
broken time reversal symmetry of the pair potential.
Here, we consider the contribution of bound states to $\sigma_T (eV)$
in the $F/I/S$ junctions
in relation to the wave function of the Andreev reflected
hole and the critical angle
$\theta_{C}$ for $0<eV<\Delta_{0}$.
The magnitude of $\sigma_{T}(eV)$ is determined
based on the magnitude of the Andreev reflection coefficient
through the energy level of the bound state.
For $\theta_M =0$ or $\pi$, when an electron
with a majority spin is injected, the Andreev reflected hole
has a minority spin and its wave function becomes
a propagating wave for $|\theta_S |<\theta_C$,
and an evanescent wave for $|\theta_S |>\theta_C$.
At the same time, an electron injection with a minority
spin is prohibited for $|\theta_S |>\theta_C$.
Therefore, an injected electron with both majority and
minority spin does not
contribute to $\sigma_T (eV)$ through the bound state
for $|\theta_S |>\theta_C$.
Due to the vanishing of the contribution of bound states
for $|\theta_S |>\theta_C$,
the change in the curvature at
$eV=E_C$ in $\sigma_T (eV)$ occurs,
where $E_C$ is estimated based on the energy of the bound state
formed at $\theta_S =\theta_C$ (see Fig. 3(b)).
The magnitude of $\theta_C$ is reduced for larger $X$ value, and
at the same time $E_{C}$ is also reduced.
For $X=1$, the structure vanishes since $\theta_C =0$ and $E_{C}=0$.
Next, we look at the case in which $\theta_M =\pi /2$ is satisfied,
where an injected electron with a majority (minority) spin
is reflected as a hole with a majority (minority) spin.
When an electron
with majority spin is injected, the Andreev reflected hole
has a majority spin and its wave function always becomes
a propagating wave independent of $\theta_S$,
while injection of an electron with a minority
spin is prohibited for $|\theta_S |>\theta_C$ similar to the case
in which $\theta_{M}=0$ or $\pi$.
Therefore, only an injected electron with
a minority spin does not
contribute to $\sigma_T (eV)$ through the bound state
for $|\theta_S |>\theta_C$.
Due to the vanishing of the contribution of bound states
for $|\theta_S |>\theta_C$,
the change in the curvature at
$eV=E_C$ in $\sigma_T (eV)$ occurs as in the case
in which $\theta_{M}=0$ or $\pi$.
\par
Finally, we discuss the tunneling conductance for the non-unitary case
presented in Sr$_2$RuO$_4$, ${\it i.e.}$,
$\Delta_{\uparrow\uparrow}=\Delta_0 e^{i\theta_S}$, otherwise $=0$,
where the $d$-vector is parallel to the $ab$-plane
of triplet superconductors.
In this superconducting state,
only an electron injection with an up spin feels the pair potential.
In the present non-unitary case, the pair potential written in the coordinate
of spin space in the ferromagnet is given by
\begin{eqnarray}
\hat{\Delta}^{F}(\theta_S )=\left(
\begin{array}{cc}
\cos^{2}{\frac{\theta_M}{2}} & -\frac{1}{2}\sin{\theta_M}\\
-\frac{1}{2}\sin{\theta_M} & \sin^{2}{\frac{\theta_M}{2}}
\end{array}
\right) e^{-i\phi_M}\Delta_0 e^{i\theta_S}.
\end{eqnarray}
%
\par
The $\theta_M$ dependence of $\sigma_T (eV=0)$
is plotted in Fig. 4 with $Z=5$.
For $\theta_M =0$ ($\theta_M =\pi$),
since the relation
$a_{\bar{\downarrow}\bar{\downarrow}}
=a_{\bar{\uparrow}\bar{\downarrow}}
=a_{\bar{\downarrow}\bar{\uparrow}}=0$
($a_{\bar{\uparrow}\bar{\uparrow}}
=a_{\bar{\uparrow}\bar{\downarrow}}
=a_{\bar{\downarrow}\bar{\uparrow}}=0$) holds,
the Andreev reflection with only a majority (minority) spin exists
for an electron injection with a majority (minority) spin.
Therefore, $\sigma_T (eV=0)$ for $\theta_M =0$ is larger than
that for $\theta_M =\pi$.
It is a significant fact that
$\sigma_{T}(eV=0)$ has a different period as a function of $\theta_M$
as compared to that in the unitary pairing.
Using these properties, we can distinguish
the unitary and non-unitary states
which are presented as a promising symmetry of Sr$_2$RuO$_4$.
\par
In conclusion, we have studied tunneling conductance in $F/I/S$ junctions
by changing the angle $\theta_M$
between the direction of the magnetization axis of
the ferromagnet and the $c$-axis of the superconductor.
The $\theta_M$ dependence appears only
in the triplet superconducting junctions and
its dependence is sensitive to the
direction of the $d$-vector in triplet superconductors.
From these properties, tunneling spectroscopy of
$F/I/S$ junctions for various directions of the
magnetization axis becomes
a powerful method for identifying the parity of the superconductor.
Moreover we can identify detailed profiles of
the triplet superconducting pair potentials.
In the present paper, we neglected the spin-orbit coupling
in triplet superconductors.
If the magnitude of spin-orbit coupling is not strong,
we expect the obtained results will not be changed qualitatively.
\par
In actual experiments, we propose a
SrRuO$_3$ / Sr$_2$RuO$_4$ junction as
a promising candidate for a $F/I/S$ junction,
where SrRuO$_3$ is known as a ferromagnet
with a Curie temperature $T_{Q}$ much higher than the
transition temperature of Sr$_2$RuO$_4$ ($T_{C}$).
The direction of the magnetization axis of SrRuO$_3$
is fixed at $T_{C}<T<T_Q$, and by decreasing the temperature
below $T_{C}$ we can realize $F/I/S$ junctions with an arbitrary
direction of the magnetization axis.
The response of conductance spectra discussed in the present paper
is easily accessible in actual experimental situations.
We hope that the present theoretical predictions
will be validated in future experiments.
\par
This work was partially supported by the Core Research for
Evolutional Science and Technology (CREST) of the Japan Science and
Technology Corporation (JST).


\newpage
\noindent
Figure Captions \par
\noindent
Fig. 1.
Schematic illustration of
ferromagnet / superconductor junction.
The direction of the magnetization axis is denoted by
a polar coordinate ($\theta_M$,$\phi_M$).
\par
\vspace{12pt}
\noindent
Fig. 2.
(a) Energy dependence and 
(b) $\theta_M$ dependence
of the normalized conductance
for unitary symmetry.
In Fig. 2(a),
$a$: $\theta_M =0$,
$b$: $\theta_M =\pi /4$ and $c$: $\theta_M =\pi /2$ with
$Z=0$, $X=0.999$, and
$d$: $Z=0$ with $X=0$.
In Fig. 2(b),
$a$: $Z=0$, $b$: $Z=1$ and $c$: $Z=5$ with $X=0.9$, $eV=0$.
$d$: The $d_{x^2 -y^2}$ state for $Z=1$ with $X=0.9$
in the case of the existence of ZES.
\par
\vspace{12pt}
\noindent
Fig. 3.
Energy dependence of the normalized conductance
for unitary symmetry.
$a$: $\theta_M =0$,
$b$: $\theta_M =\pi /2$ for $X=0.9$, and
$c$: $\theta_M =0$,
$d$: $\theta_M =\pi /2$ for $X=0.999$ with $Z=10$.
The inset shows 
$a$: bound states formed at the surface of the triplet superconductor
in the unitary case, and
$b$: bound states for $0\leq \theta_S <\theta_C$
in the ferromagnet / insulator / triplet superconductor junction
with $X=0.9$.
\par
\vspace{12pt}
\noindent
Fig. 4
$\theta_M$ dependence of the normalized conductance
for non-unitary symmetry.
$a$: $X=0.7$, $b$: $X=0.9$ and $c$: $X=0.999$ with $Z=5$, $eV=0$.
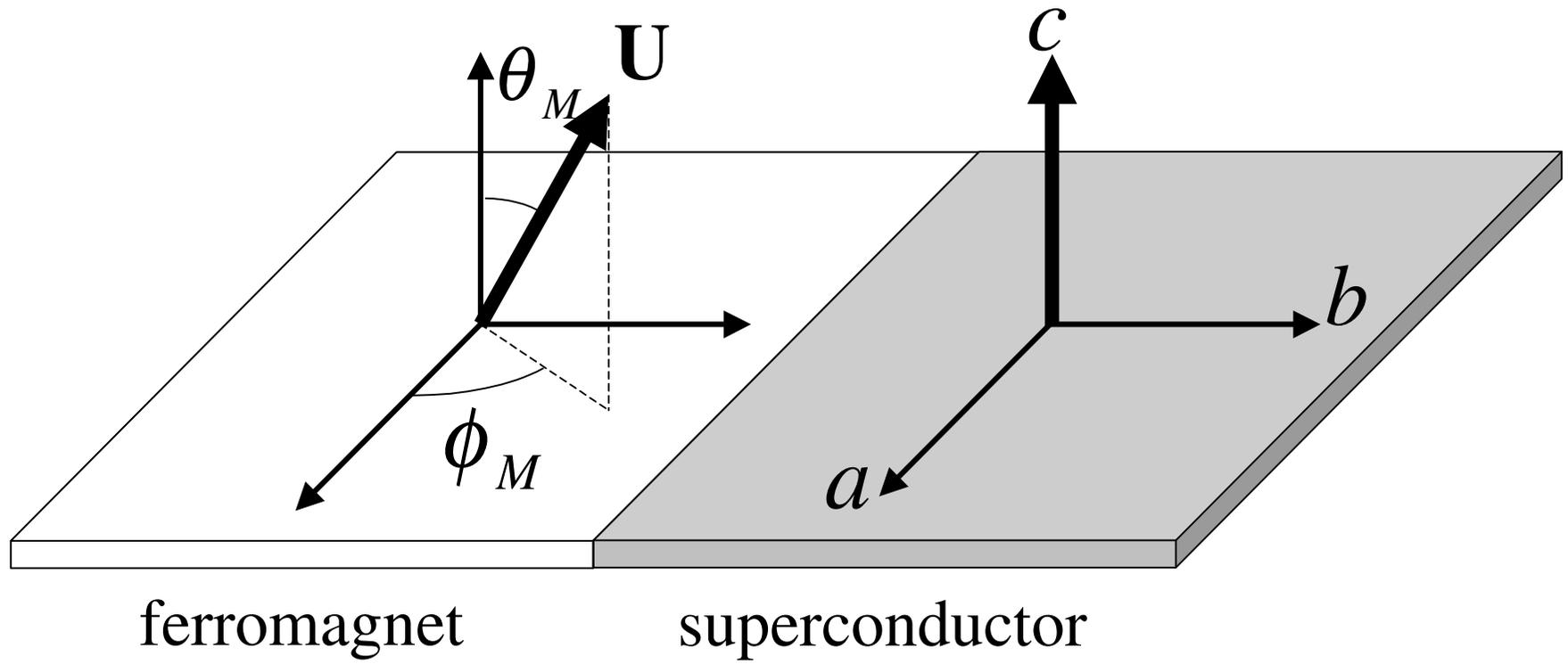
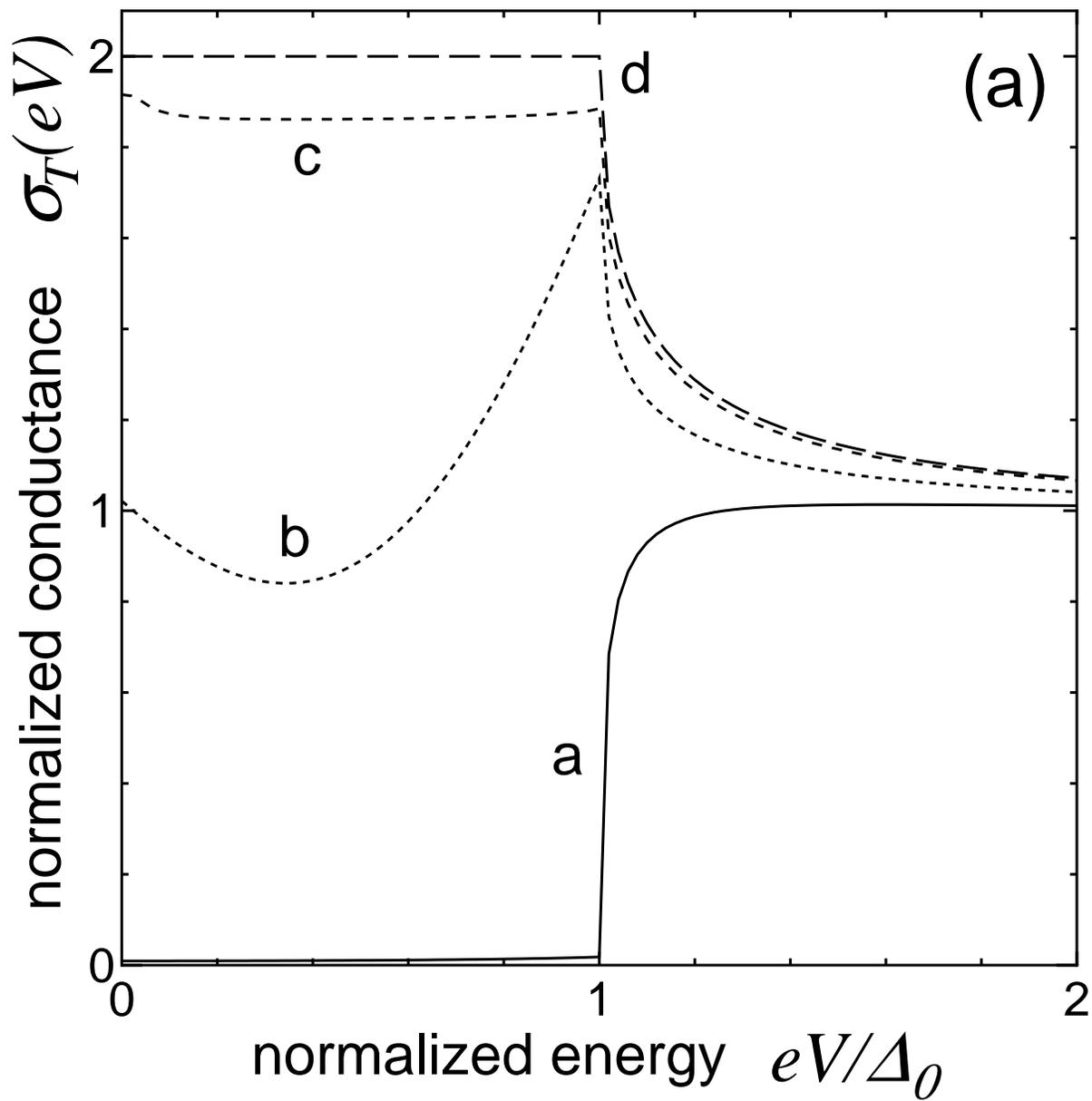
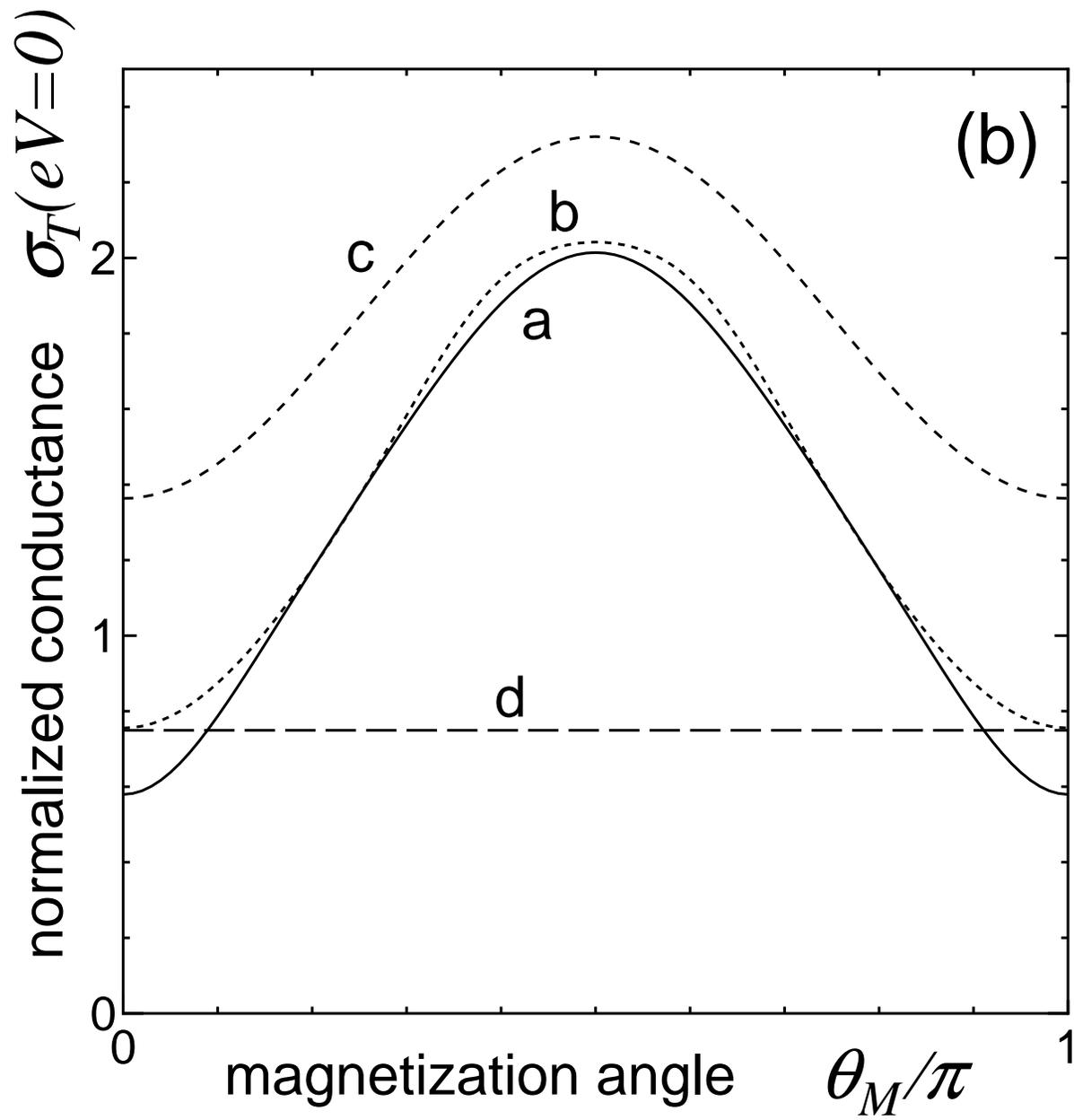
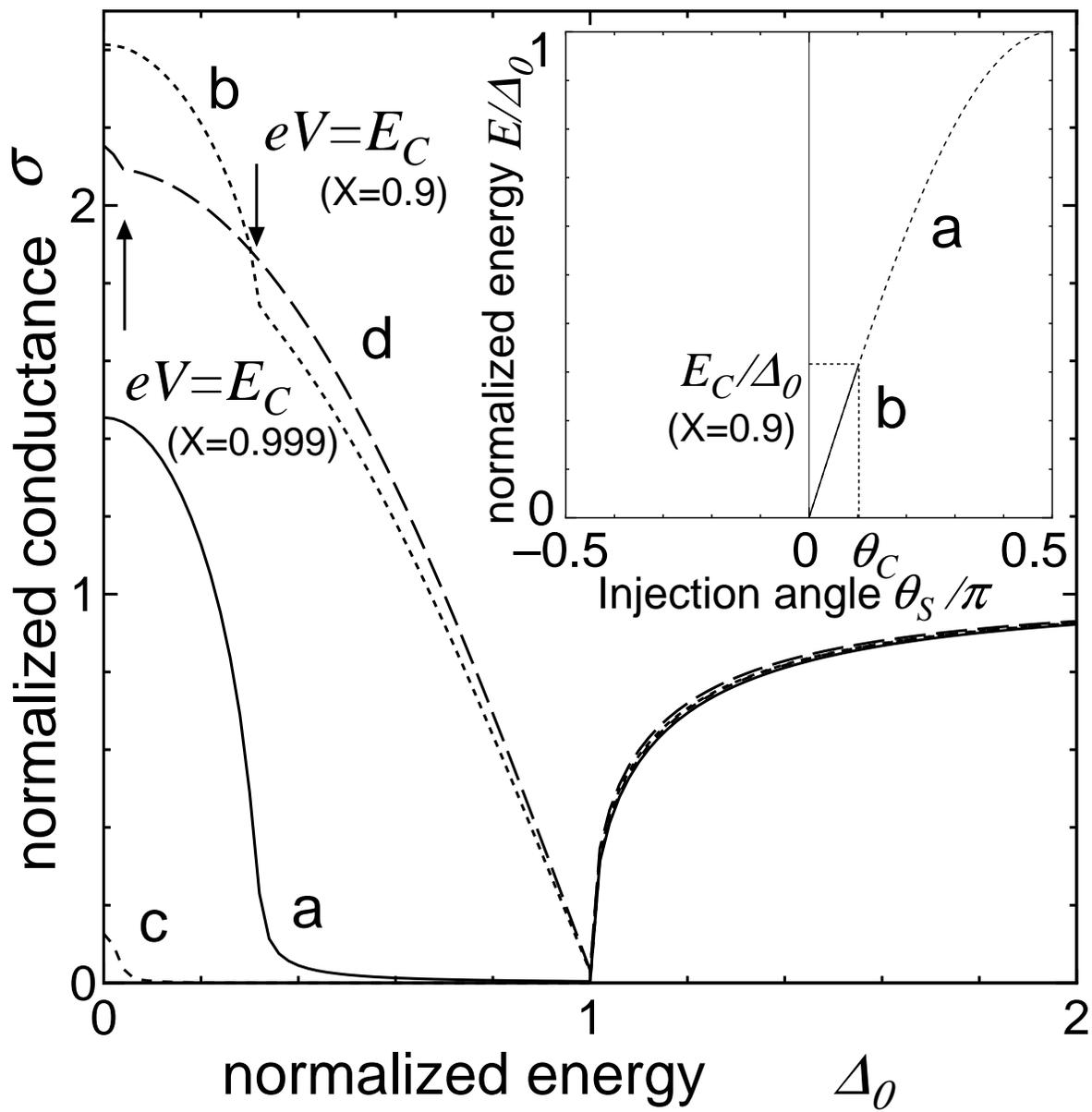
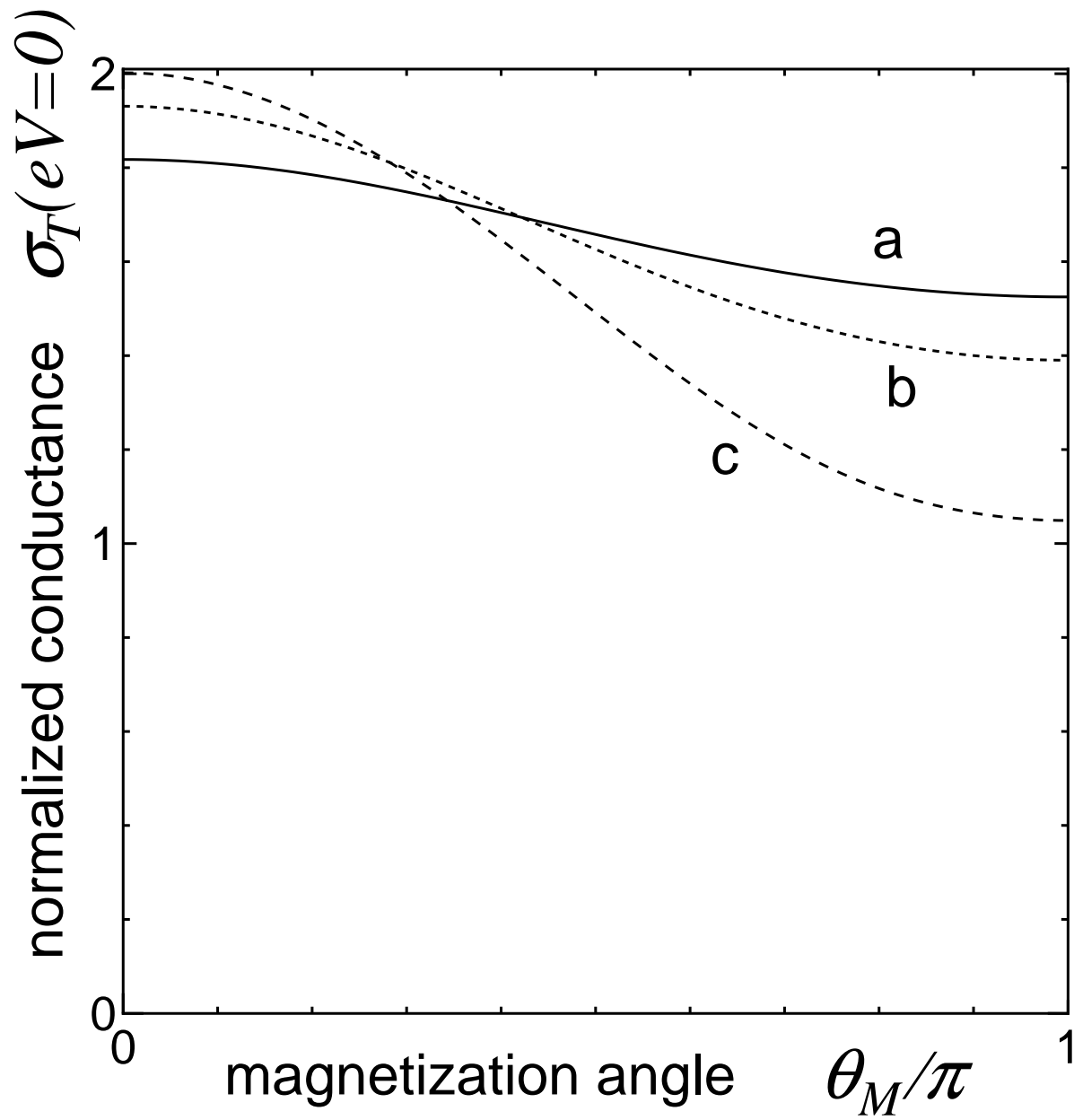
\end{document}